\documentclass[11pt,a4paper]{article}

\pdfoutput=1
\usepackage{jheppub}
\usepackage[normalem]{ulem}
\usepackage{slashed}
\usepackage{amsmath}
\def\beq{\begin{equation}}
\def\eeq{\end{equation}}
\def\bea{\begin{eqnarray}}
\def\eea{\end{eqnarray}}
\def\nn{\nonumber}

\def\roughly#1{\mathrel{\raise.3ex\hbox
{$#1$\kern-.75em\lower1ex\hbox{$\sim$}}}}

\def\gsim{\roughly>}

\def\sla#1{\raise.15ex\hbox{$/$}\kern-.57em #1}% Feynman slash

\usepackage{physics} % Physics shortcuts
\usepackage{booktabs} % Nice rules for tables: toprule, midrule, bottomrule

\begin{document}

\title{LHC Constraints on Scalar Diquarks}

\author[a]{Bruna Pascual-Dias,}
\author[b]{Pratishruti Saha}
\author[a]{and David London}
\affiliation[a]{Physique des Particules, Universit\'e de Montr\'eal, \\
C.P. 6128, succ.\ centre-ville, Montr\'eal, QC, Canada H3C 3J7}
\affiliation[b]{Harish-Chandra Research Institute, Chhatnag Road, \\
Jhunsi, Allahabad - 211019, India}
\emailAdd{bruna.pascual.dias@umontreal.ca}
\emailAdd{pratishrutisaha@hri.res.in}
\emailAdd{london@lps.umontreal.ca}

\abstract{A number of years ago, low-energy constraints on scalar
  diquarks, particles that couple to two quarks, were examined. It was
  found that the two most weakly-constrained diquarks are $D^u$ and
  $D^d$, colour antitriplets that couple to $u_{R}^{i} u_{R}^{j}$ and
  $d_{R}^{i} d_{R}^{j}$, respectively. These diquarks have not been
  observed at the LHC. In this paper, we add the LHC measurements to
  the low-energy analysis, and find that the constraints are
  significantly improved. As an example, denoting $x^u$ as the $D^u$
  coupling to the first and second generations, for $M_{D^u} = 600$ GeV,
  the low-energy constraint is $|x^u| \leq 14.4$, while the addition
  of the LHC dijet measurement leads to $|x^u| \leq 0.13$--$0.15$.
  Further improvements are obtained by adding the measurement of
  single top production with a $p_T$ cut. These new constraints must
  be taken into account in making predictions for other low-energy
  indirect effects of diquarks.}


%\arxivnumber{}

\preprint{
{\flushright
UdeM-GPP-TH-20-279 \\
}}

\maketitle

\section{Introduction}

The Standard Model (SM) has been extremely successful in explaining
almost all experimental measurements to date. However, for a variety
of reasons -- the hierarchy problem, dark matter, CP violation and the
matter-antimatter asymmetry, etc.\ -- it is generally believed that it
is not complete. There must be physics beyond the SM. It was hoped
that the LHC would produce new-physics (NP) particles directly, but so
far this has unfortunately not happened. The scale of NP may be above
the present reach of the LHC. 

Still, even if this is the case, all hope is not lost: one can also
search for NP through indirect signals. (Indeed, there are currently
indirect hints of NP in $b \to s \mu^+\mu^-$ and $b\to c \tau^-
{\bar\nu}_\tau$ transitions \cite{London:2019nlu}.) Of course, for a
particular kind of NP, if one wants to examine how large the indirect
effects can be in a given process, one must include the constraints on
its mass and couplings derived from direct searches.

One possible type of NP is a diquark, a particle that couples to two
quarks. A diquark can be a scalar or a vector, and transforms as a
${\mathbf 6}$ or ${\mathbf{\bar 3}}$ of $SU(3)_C$. In this paper, we
focus on scalar diquarks. These appear in models with $E_6$
\cite{Hewett:1988xc} or $SU(2)_L \times SU(2)_R \times SU(4)_C$
\cite{Mohapatra:2007af} symmetry, and in supersymmetry with R-parity
violation \cite{Barbier:2004ez}. Studies of diquark phenomenology
mostly fall into three categories: (i) the LHC discovery reach for
scalar diquarks \cite{Mohapatra:2007af, Tanaka:1991nr, Atag:1998xq,
  Cakir:2005iw, Chen:2008hh, Han:2009ya, Gogoladze:2010xd,
  Berger:2010fy, Han:2010rf, Baldes:2011mh, Richardson:2011df,
  Karabacak:2012rn, Kohda:2012sr, Chivukula:2015zma, Zhan:2013sza,
  Liu:2013hpa}, (ii) explanations of the Tevatron $t{\bar t}$
forward-backward asymmetry \cite{Shu:2009xf, Dorsner:2009mq,
  Dorsner:2010cu, Arhrib:2009hu, Ligeti:2011vt, Hagiwara:2012gy,
  Allanach:2012tc, Dupuis:2012is, Han:2012dd}, and (iii) contributions
to $n$-${\bar n}$ oscillations \cite{Baldes:2011mh, Mohapatra:1980qe,
  Babu:2008rq, Ajaib:2009fq, Gu:2011ff, Babu:2012vc, Arnold:2012sd,
  Babu:2013yca}. Thus, there is a good deal of room to examine the
effect of diquarks in other processes (for example, see
Refs.~\cite{Beaudry:2017gtw, Chen:2018stt, Dev:2020qet}).

In 2011, Giudice, Gripaios and Sundrum (GGS) \cite{GGS} considered
diquarks with sizeable flavour-dependent couplings to light quarks,
and examined the low-energy constraints\footnote{Other analyses have
  examined the constraints on diquarks from LEP data
  \cite{Bhattacharyya:1995bw} and flavour physics
  \cite{Fortes:2013dba}.} from flavour-changing neutral currents,
electric-dipole moments and neutral meson mixing. This was done for
both the ${\mathbf 6}$ and ${\mathbf{\bar 3}}$ diquarks. They found
that two types of diquark, both transforming as a ${\mathbf{\bar 3}}$
under $SU(3)_C$, were rather immune to the constraints. That is, they
could be rather light even while keeping reasonably large couplings.
They are $D^u$ and $D^d$, diquarks that couple to $u_{R}^{i}
u_{R}^{j}$ and $d_{R}^{i} d_{R}^{j}$, respectively (here $i$ and $j$
are generation indices).  They encouraged the search for these scalar
diquarks at the LHC.

Now, we know that, to date, diquarks have not been observed at the
LHC. But this does not exclude the possibility of measurable indirect
effects in low-energy processes. As noted above, if one wants to
predict how large such effects can be in a particular process, it is
important to take into account the constraints from direct searches on
the diquark's mass and couplings.

With this in mind, in this paper we extend the GGS analysis to include
the constraints from the LHC. These come in two types. First, there
are the constraints from direct searches, which apply to both $D^u$
and $D^d$. Second, there are indirect constraints on $D^u$ due to its
contribution to top-quark production. Processes that can potentially
be important include the production of $t {\bar t}$, $t t$ and single
top production. We will show that the LHC constraints reduce the
allowed parameter space of diquark masses and couplings compared to
GGS.

We begin in Sec.~\ref{sec:scalar_diquarks} with a summary of the various
scalar diquarks. The low-energy (GGS) constraints are reviewed in 
Sec.~\ref{sec:low_energy_constraints}. Sec.~\ref{sec:direct_searches} 
contains our analysis of the constraints from direct searches at the LHC.
Constraints from single top production are examined in Sec.~\ref{sec:stp}. 
We conclude in Sec.~\ref{sec:conclusions}.

\section{Scalar Diquarks}
\label{sec:scalar_diquarks}

We consider the addition of a scalar diquark to the SM. This scalar
diquark $D$ has mass $M_D$, spin 0, and couples to a pair of quarks.
Similar to the parametrization of Ref. \cite{Han:2009ya}, we write the
interaction Lagrangian after electroweak symmetry breaking as
\beq
\mathcal{L} = 
  % \;2\sqrt{2}
  \;\sqrt{2}
  \;{\overline{K}_k}^{ab}
  \;D^{k}
  \;\overline{q^{i}_{a}}
  \;\lambda_{ij}
  \;P_{L,R}
  \;{q^{j}_{b}}^{C}
  + \text{h.c.}
\eeq
Here $a,b \in \{1,2,3\}$ are colour indices, $i,j \in \{1,2,3\}$ are
generation indices, $P_{L,R} \equiv (1 \mp \gamma^{5})/2$ is the left-
or right-chirality projection operator, and $q^{C} \equiv
C\;{\overline{q}}\,^{T}$ is the conjugate quark field. Since it
couples to two quarks, the diquark $D^k$ transforms as a ${\mathbf 6}$
or ${\mathbf{\bar 3}}$ of $SU(3)_C$; the index $k$ runs over the
components of the representation (1 to 6 for a ${\mathbf 6}$, 1 to 3
for a ${\mathbf{\bar 3}}$). The ${\overline{K}_k}^{ab}$ are the
$SU(3)_C$ Clebsch-Gordan coefficients coupling this representation to
two ${\mathbf 3}$s. $\lambda_{ij}$ is the coupling to the $i$ and $j$
generations. Note that the two quarks coupling to the
diquark have the same chirality.

The Clebsch-Gordan coefficients ${K^{k}}_{ab}$ for the ${\mathbf 6}$
diquark representation are symmetric, while for the ${\mathbf{\bar
    3}}$ diquark representation the ${K^{k}}_{ab}$ are antisymmetric.
Given that a ${\mathbf{\bar 3}}$ diquark is an antifundamental
representation of $SU(3)_C$ \cite{Han:2009ya}, we can assign it a
single colour index~$c$. This allows us to write the Lagrangian for a
${\mathbf{\bar 3}}$ diquark as
\beq
  \mathcal{L}^{\mathbf{\bar{3}}} = 
  % 2
  \;\epsilon^{cab}
  % \;\overline{D_{k}}
  \;D_{c}
  \;\overline{q^{i}_{a}}
  \;\lambda_{ij}
  \;P_{L,R}
  \;{q^{j}_{b}}^{C}
  + \text{h.c.}.
\eeq

When two quarks combine to form a diquark, the symmetry of the
combined state is directly dependent on the individual symmetries
under $SU(3)_C$ and $SU(2)_L$. For instance, consider $Q_L Q_L$. Under
$SU(3)_C$, the combination of two ${\mathbf 3}$s produces a ${\mathbf
  6}$ (symmetric) and a ${\mathbf{\bar 3}}$ (antisymmetric). Under
$SU(2)_L$, ${\mathbf 2} \times {\mathbf 2}$ yields a ${\mathbf 3}$
(symmetric) and a ${\mathbf{1}}$ (antisymmetric). Thus, the state
$(\mathbf{\bar{3}}, \mathbf{1})_{+1/3}$ of $SU(3)_C \times SU(2)_L$ is
overall symmetric. Similarly, if one considers $u_R u_R$, the state
$(\mathbf{\bar{3}}, \mathbf{1})_{+4/3}$ is antisymmetric.

The symmetry of the combined state under interchange of quarks has
important implications for the couplings of the diquark. If the
diquark state is antisymmetric under the exchange of two quarks, then
the coupling to two quarks of the same flavour vanishes. As a
consequence, we see that the antisymmetric diquarks only couple to
pairs of quarks with different flavours.  We thus see that the
antisymmetry of the diquark state under the SM gauge group entails an
antisymmetry under flavour \cite{GGS}. This antisymmetry has important
consequences, as it implies that any flavour-changing diagram must
involve all three generations of quarks \cite{GGS}.

With this in mind, all possible scalar diquarks that couple to quarks
within the SM can be classified by their charges under the SM gauge
group $SU(3)_C \times SU(2)_L \times U(1)_Y$ \cite{Han:2010rf,GGS}.
They are presented in Table~\ref{tab:diquarks}, where we use the
convention $Q = I_3 + Y$.

\begin{table}[!t]
\centering
\begin{tabular}{ccccc}
\toprule
Name & $SU(3)_C$ & $SU(2)_L$ & $U(1)_Y$ & Coupling  \\
\midrule
I & ${\mathbf 6}$       & ${\mathbf 3}$       & $+1/3$   & ($Q_L Q_L$) \\
II & ${\mathbf{\bar{3}}}$ & ${\mathbf 3}$       & $+1/3$   & [$Q_L Q_L$] \\
III & ${\mathbf 6}$       & ${\mathbf 1}$       & $+1/3$   & [$Q_L Q_L$], $u_R d_R$ \\
IV & ${\mathbf{\bar{3}}}$ & ${\mathbf 1}$       & $+1/3$   & ($Q_L Q_L$), $u_R d_R$ \\
V & ${\mathbf 6}$       & ${\mathbf 1}$       & $+4/3$   & ($u_R u_R$) \\
VI $\equiv D^u$ & ${\mathbf{\bar{3}}}$ & ${\mathbf 1}$       & $+4/3$   & [$u_R u_R$] \\
VII & ${\mathbf 6}$       & ${\mathbf 1}$       & $-2/3$   & ($d_R d_R$) \\
VIII $\equiv D^d$ & ${\mathbf{\bar{3}}}$ & ${\mathbf 1}$       & $-2/3$   & [$d_R d_R$] \\
\bottomrule
\end{tabular}
\caption{Scalar diquarks classified by their charges under the SM
  gauge group. In the `Coupling' column, parentheses indicate a
  symmetric coupling and square brackets indicate an antisymmetric
  coupling with respect to flavour indices \cite{GGS}.}
\label{tab:diquarks}
\end{table}

\section{Low-energy Constraints}
\label{sec:low_energy_constraints}

In Ref.~\cite{GGS}, GGS worked out the low-energy constraints on these
diquarks. We review their results in this section.

First, if one imposes only the $SU(3)_C \times SU(2)_L \times U(1)_Y$
gauge symmetry, diquarks can also have a dimension-four Yukawa-type
coupling to a lepton and a quark \cite{Assad:2017iib}. The presence of
both diquark and leptoquark couplings would lead to proton decay
\cite{GGS}, and would of course place extremely stringent constraints
on the diquarks' couplings and masses. This would essentially rule out
any effects at the TeV scale and below. In order to avoid this, there
must be an additional global symmetry, such as lepton number or baryon
number, that forbids this leptoquark coupling.

\begin{figure}[!t]
\centering
\includegraphics[width=1.0\textwidth]{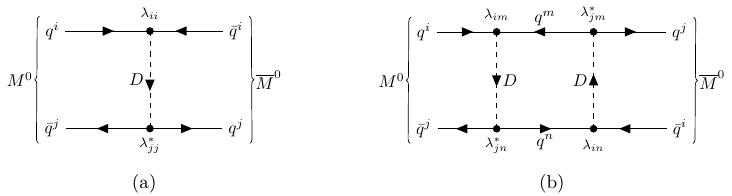}
\caption{Diquark contributions to $M^0$-${\bar M}^0$ mixing at (a) tree
  level and (b) one loop.}
\label{fig:Mmixing}
\end{figure}

Diquarks I, V and VII all contribute at tree level to the $\Delta F =
2$ process $M^0$-${\bar M}^0$ mixing ($M = K$, $D$, $B_d$, $B_s$)
(see Fig.~\ref{fig:Mmixing}\hyperref[fig:Mmixing]{a}). This leads to
very strong constraints on these diquarks. The other diquarks
contribute to meson mixing at one loop via a box diagram (see
Fig.~\ref{fig:Mmixing}\hyperref[fig:Mmixing]{b}). However, not all
contributions are the same size. Consider diquark II. Since it couples
only to left-handed quarks, there is also a box diagram in which one
of the internal diquarks is replaced by a $W$. And since the $W$ is
considerably lighter than the $D$, this amplitude is larger than the
analogous amplitude with two virtual diquarks, leading to stronger
constraints on diquark II. As for diquarks III and IV, they couple to
both left- and right-handed $u^i d^j$ pairs. This leads to non-chiral
$\Delta F = 2$ operators which are greatly enhanced when compared to
the chiral operators. Once again, this leads to stronger constraints
on diquarks III and IV. This result holds even in the case where one
of the couplings (left- or right-handed) dominates \cite{GGS}.

The upshot is that, of the eight possible diquarks, two of them -- 
VI~$\equiv~D^u$ and VIII~$\equiv~D^d$ -- are more weakly constrained than
the others. It is for this reason that GGS suggested that these scalar
diquarks be searched for at the LHC. In this paper, we examine the
additional constraints on these diquarks from the LHC using direct
searches and measurements of top production.

The $D^q$ diquark ($q = u, d$) couples to $q_{R}^{i} q_{R}^{j}$. Since it
transforms as $(\mathbf{\bar{3}}, \mathbf{1})$ under $SU(3)_C \times
SU(2)_L$, the coupling $\lambda_{ij}$ is antisymmetric.  In their
analysis, GGS write this coupling as
\beq 
\lambda^q_{ij} \equiv \epsilon_{ijk} \lambda^q_{k}
  % \quad\Rightarrow\quad
  % \begin{cases}
  % \lambda_1 \equiv \lambda_{23} = -\lambda_{32}\\
  % \lambda_2 \equiv \lambda_{31} = -\lambda_{13}\\
  % \lambda_3 \equiv \lambda_{12} = -\lambda_{21}\\
  % \end{cases}
  \quad\Longrightarrow\quad
  \lambda^q = 
  \begin{pmatrix}
  0 & \lambda^q_3 & -\lambda^q_2\\
  -\lambda^q_3 & 0 & \lambda^q_1\\
  \lambda^q_2 & -\lambda^q_1 & 0\\
  \end{pmatrix} ~.
\eeq
The constraints on the $\lambda^q_i$ come from a variety of
processes. For real couplings, which we consider in our analysis,
they are as follows. For $\lambda^d_i$, they include $K^0$-${\bar
  K}^0$ and $B_d^0$-${\bar B}_d^0$ mixing, $b \to s\gamma$ and $b \to
d\gamma$, $R_b$, and $B^\pm \to \phi\pi^\pm$. For $\lambda^u_i$, there
are only $D^0$-${\bar D}^0$ mixing and $A_c$ (defined in terms of the
coupling of the $Z$ boson to charm quarks). The constraints from these
various quantities are given in Table~\ref{tab:constraints}.

\begin{table}[!t]
\centering
\begin{tabular}{cc}
\toprule
Process & Bound ($M_D$/TeV) \\
\midrule
$\Delta m_K$ & $|\lambda_1^d \lambda_2^d| \le 4.6 \times 10^{-2}$ \\
$B_d^0$-${\bar B}_d^0$ mixing & $|\lambda_1^d \lambda_3^d| \le 3.6 \times 10^{-2}$ \\
$b \to s\gamma$ & $\sqrt{|\lambda_2^d \lambda_3^d|} \le 1.8$ \\
$b \to d\gamma$ & $\sqrt{|\lambda_1^d \lambda_3^d|} \le 0.9$ \\
$R_b$ & $|\lambda_{1,2}^d| \le 24$ \\
$B^\pm \to \phi\pi^\pm$ & $\sqrt{|\lambda_1^d \lambda_3^d|} \le 0.1$ \\
\midrule
$D^0$-${\bar D}^0$ mixing & $|\lambda_1^u \lambda_2^u| \le 1.5 \times 10^{-2}$ \\
$A_c$ & $|\lambda_3^u| \le 24$ \\
\bottomrule
\end{tabular}
\caption{Bounds in units of $M_D$/TeV on the couplings of the diquarks
  $D^d$ and $D^u$ \cite{GGS}.}
\label{tab:constraints}
\end{table}

In our analysis, we distinguish between the couplings that involve
only light quarks (first and second generations) and those involving
the third generation of quarks (we assume the couplings involving the
first and third generations have the same magnitude as those involving
the second and third generations). In the GGS convention, this
corresponds to setting $\lambda_1^q = \lambda_2^q = y^q$ and
$\lambda_3^q = x^q$.  The coupling matrix is then given by
\beq
  \lambda^q =
  \begin{pmatrix}
  0 & x^q & -y^q\\
  -x^q & 0 & y^q\\
  y^q & -y^q & 0\\
  \end{pmatrix} ~.
\eeq

We can now translate the GGS bounds to our notation. For $D^u$, the
bound from $D^0$-${\bar D}^0$ mixing yields $\abs{y^u}^2 \leq 1.5
\times 10^{-2} \;(M_D/{\rm TeV})$, or $\abs{y^u} \leq 0.12
\;\sqrt{(M_D/{\rm TeV})}$. The bound from the electroweak precision
tests ($A_c$) is $\abs{x^u} \le 24 \;(M_D/{\rm TeV})$. For $D^d$, the
$K^0$-${\bar K}^0$ mass difference $\Delta m_K$ imposes that
$|y^d| \leq \sqrt{4.6 \times 10^{-2} (M_D/\text{TeV})}$. Similarly, the
constraints from $B_d^0$-$\bar{B}_d^0$ mixing require $|x^d y^d|
\leq 3.6 \times 10^{-2} (M_D/\text{TeV})$.

In our analysis, we consider two diquark masses, $M_D = 600$ GeV and
$M_D = 1$ TeV. For these two masses, the constraints are\footnote{For
  $|x^d y^d|$, we consider the constraint from $B_d^0$-${\bar B}_d^0$
  mixing, as the constraint from $B^\pm \to \phi\pi^\pm$, though
  apparently more stringent, involves additional theoretical
  assumptions \cite{GGS}.}
\begin{align}
\label{constraints}
M_D & = 600~{\rm GeV} & M_D & = 1~{\rm TeV} \nn\\
& |y^u| \leq 0.09 ~, & & |y^u| \leq 0.12 ~, \nn\\
& |x^u| \leq 14.4 ~, & & |x^u| \leq 24 ~, \\
& |y^d| \leq 0.17 ~, & & |y^d| \leq 0.21 ~, \nn\\
& |x^d y^d| \leq 0.022 ~, & & |x^d y^d| \leq 0.036 ~. \nn
\end{align}

\section{LHC Constraints: Direct Searches}
\label{sec:direct_searches}

\begin{figure}[!t]
\centering
\includegraphics[width=1\textwidth]{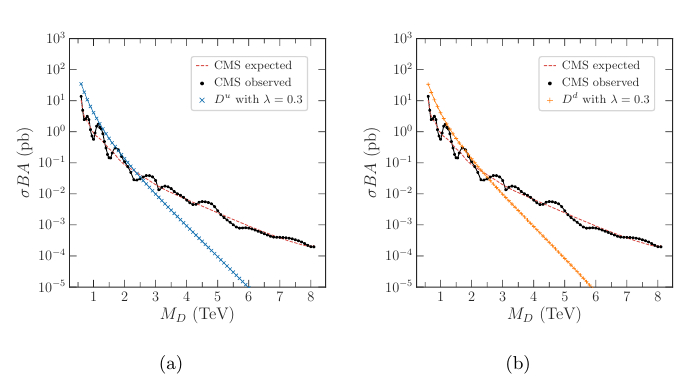}
\caption{Theoretical predictions of $\sigma B A$ for diquarks $D^u$
  (blue cross) and $D^d$ (orange plus sign), for the case where all
  couplings are equal and of electromagnetic strength.  Observed
  values at 95\%~CL (black dots) and expected values (red dashed line)
  from Ref.~\cite{Sirunyan:2018xlo} are presented for comparison.}
\label{fig:sBAvsM}
\end{figure}

In this section, we obtain constraints on the diquark parameter space
using measurements of dijet production at the LHC.  In
Ref.~\cite{Sirunyan:2018xlo}, the CMS Collaboration presents
measurements of narrow dijet resonances at $\sqrt{s} = 13$ TeV.  It is
found that the data exclude the scalar diquarks in $E_6$ models
\cite{Hewett:1988xc} with a coupling constant of electromagnetic
strength for masses less than $7.2$ TeV. This result is obtained from
the model-independent observed 95\% CL upper limits on the product
$\sigma B A$ for quark-quark resonances. Here $\sigma$ is the
production cross section, $B$ is the branching ratio for a dijet
decay, and $A$ is the acceptance, which includes kinematic
requirements of the dijet final state.  The observed and expected
values of $\sigma B A$ from Ref.~\cite{Sirunyan:2018xlo} are presented
in Fig.~\ref{fig:sBAvsM} as black dots and a red dashed line,
respectively.  Ref.~\cite{Sirunyan:2018xlo} states that these limits
can be directly compared to parton-level calculations of $\sigma B A$
without detector simulation.

In order to get constraints from direct searches, we compare these
results with the predictions of $\sigma B A$ for the scalar diquarks
$D^u$ and $D^d$. These are obtained by calculating the leading-order
(LO) cross section for the production process $pp \to D^{u,d}$, as
well as the branching ratio for the decay of the corresponding diquark
into light quarks, namely $BR(D^u \to u\,c)$ and $BR(D^d \to
d\,s)$. The acceptance is calculated following the prescription in
Ref.~\cite{Sirunyan:2018xlo}. It is defined as $A = A_{\Delta}
A_{\eta}$, where $A_{\Delta}$ is the acceptance of requiring
$\abs{\Delta\eta} < 1.3$ for the dijet system and $A_{\eta}$ is the
acceptance of also requiring $\abs{\eta} < 2.5$ for each of the
jets. Since we are considering scalar diquarks, which have isotropic
decays, we set $A_{\Delta} = 0.57$ for all masses. For diquark masses
less than $1.6$ TeV, we set $A_{\eta} = 0.95$ to account for the
decrease of acceptance in this lower mass range. For larger masses,
$A_{\eta}$ is set to $1$. In summary, the acceptance for the low-mass
range is set to $A = 0.54$ while for higher masses it is set to $A =
0.57$.

The calculations are performed using MadGraph5\_amc@NLO version 2.7.2
\cite{Alwall:2014hca} by implementing the NP alongside the SM with
FeynRules \cite{Alloul:2013bka}. We base our implementation on the
existing model file for triplet diquarks in the FeynRules model
database \cite{TripletDiquarks}. The parton distribution function
(PDF) used is CTEQ6L1 \cite{CTEQ6L1} and both the renormalization and
factorization scales are set to the diquark mass $M_D$.  The LO cross
section calculated by MadGraph5\_amc@NLO is multiplied by an
approximate NLO $K$ factor ($K$ = 1.3) based on the results of
Ref.~\cite{Han:2009ya}.  In all calculations, the requirements for the
narrow-width approximation are satisfied: the $\Gamma/M_{D}$ ratio is
smaller than 2\% for couplings up to $0.2$ and it is smaller than 5\%
for couplings up to $0.3$.

We begin by considering the case where all the diquark couplings are
equal and of electromagnetic strength, $\lambda_{ij} = 0.3$, for
masses ranging from 600 GeV to 8.1 TeV. The predicted values of
$\sigma B A$ for both diquarks are presented in Fig.~\ref{fig:sBAvsM}.
We find the lower bounds on the masses to be $M_{D^u}, M_{D^d} \gsim
2.5$ TeV. The limits are
nearly identical for $D^u$ and $D^d$.  This is expected. For each
diquark, the production is dominated by one subprocess: $u\,c \to D^u$
and $d\,s \to D^d$. At the same time, the relative densities of the
initial-state quarks inside the proton roughly follow the order $u > d
> s > c$.  The combination of the two effects means that the
predictions for $\sigma B A$ are very similar in the two cases. The
differences become more pronounced as we go to higher diquark masses.

Next, we explore the scenario where the light-quark couplings $x^q$
and the third-generation couplings $y^q$ are different. We consider
two representative diquark masses of 600 GeV and 1 TeV, and perform a
scan over the pair of couplings $(x^q,y^q)$, in which each coupling
varies from $0.05$ to $0.3$ in steps of $0.01$. For each pair, we
calculate the corresponding prediction for $\sigma B A$.  A cubic
interpolation is then performed in order to span the entire region of
interest in the parameter space. The resulting values are compared to
the observed upper limits obtained by CMS. It must be noted that, for
coupling values lower than $0.05$, the calculated partial width of the
diquark is smaller than the QCD scale, meaning that a perturbative
approach is no longer valid. For this reason we do not consider
couplings less than 0.05.

\begin{figure}[!t]
\centering
\includegraphics[width=\textwidth]{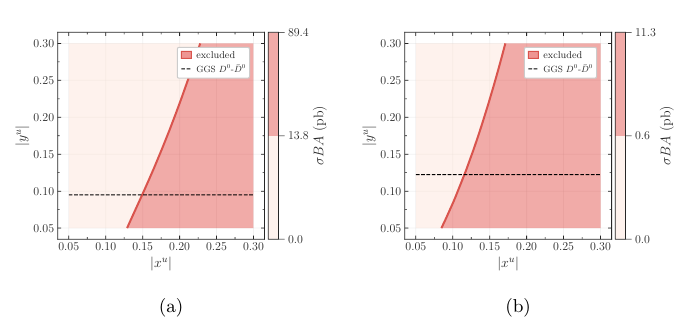}
\caption{Theoretical predictions of $\sigma B A$ for the $D^u$ diquark
  with (a) $M_{D^u} = 600$~GeV and (b) $M_{D^u} = 1$~TeV for
  different values of the light-quark coupling $|x^u|$ and the
  third-generation coupling $|y^u|$. The observed 95\%~CL upper limit on
  $\sigma B A$ \cite{Sirunyan:2018xlo} is indicated by the red solid line.
  The black dashed line denotes the GGS constraint on $|y^u|$
  [Eq.~(\ref{constraints})].}
\label{fig:dijet_uRuR}
\end{figure}

The $\sigma B A$ predictions for $D^u$ can be found in
Fig.~\ref{fig:dijet_uRuR}. We superpose the bounds from GGS
[Eq.~(\ref{constraints})] (black dashed line). The region of parameter
space above this line is excluded. We draw the bound on $\sigma B A$
from CMS (red solid line), so the region to the right of this line
(dark shade of red) is excluded and the region to the left (light
shade of red) is allowed. The net effect is that there is a
considerable improvement over the GGS bounds.  Specifically, there are
new constraints on the coupling to lighter generations, $|x^u|$. For
$M_{D^u} = 600$ GeV, $|x^u|$ must take values less than
$0.13$--$0.15$, while, for $M_{D^u} = 1$ TeV, this upper limit is
$0.08$--$0.11$. In both cases, the value of the upper limit on $|x^u|$
depends on the value of $|y^u|$.

\begin{figure}[!t]
\centering
\includegraphics[width=\textwidth]{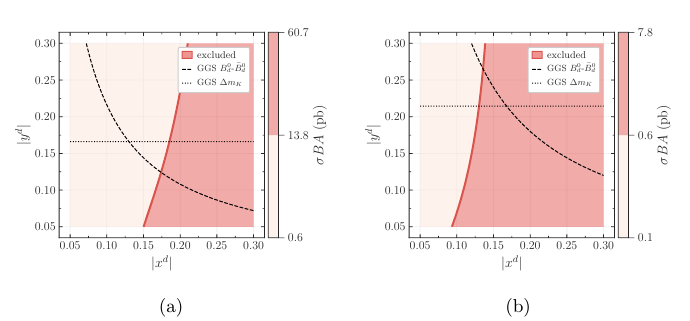}
\caption{Theoretical predictions of $\sigma B A$ for the $D^d$ diquark
  (a) $M_{D^d} = 600$~GeV and (b) $M_{D^d} = 1$~TeV, for different
  values of the light-quark coupling $|x^d|$ and the third-generation
  coupling $|y^d|$. The observed 95\%~CL upper limit on $\sigma B A$
  \cite{Sirunyan:2018xlo} is indicated by the red solid line. The black
  dashed line and black dotted line denote the GGS constraints on
  $|x^d y^d|$ and $|y^d|$, respectively [Eq.~(\ref{constraints})].}
\label{fig:dijet_dRdR}
\end{figure}

For $D^d$, the $\sigma B A$ predictions are shown in
Fig.~\ref{fig:dijet_dRdR}.  The GGS bounds [Eq.~(\ref{constraints})]
on $|x^d y^d|$ (black dashed line) and $|y^d|$ (black dotted line) are
superposed, as is the $\sigma B A$ bound from CMS (red solid line). As
above, the region to the right of this red solid line is excluded and
the region to the left is allowed. Once again, we find a significant
improvement over the GGS bounds: $|x^d| \le 0.15$--$0.17$ ($M_{D^d} =
600$ GeV) and $0.09$--$0.13$ ($M_{D^d} = 1$ TeV).

\section{LHC Constraints: Single Top Production}
\label{sec:stp}

Further LHC constraints on diquarks can come from measurements of
top-quark production. Processes potentially include the production of
$t {\bar t}$ pairs, $t t$ pairs, and single top production. These
constraints are explored in this section. Note that here we are
interested in obtaining an improvement over the constraints already
obtained from the dijet channel.

Two points are immediately obvious. First, only the $D^u$ diquark can
contribute to these processes since the $D^d$ diquark does not couple
to top quarks. Thus, any constraints apply only to $D^u$. Second,
because this diquark is an antisymmetric state, it cannot couple to
two quarks of the same flavour. As a result, it does not contribute to
$t t$ production. This leaves us with $t {\bar t}$ and single top
production.

Consider first the production of $t {\bar t}$ pairs.  At the LHC, its
cross-section is dominated by gluon-initiated processes.  The
remaining contributions arise from $q \bar{q} \to t \bar{t}$
processes, which would include contributions from the $D^u$ diquark.
However, if we restrict to couplings allowed by the GGS and dijet
data, we find that the diquark contribution is overwhelmed by the SM
contributions.  For this reason, we are not able to obtain meaningful
constraints from $t {\bar t}$ production measurements.

\begin{figure}[!t]
\centering
\includegraphics[width=\textwidth]{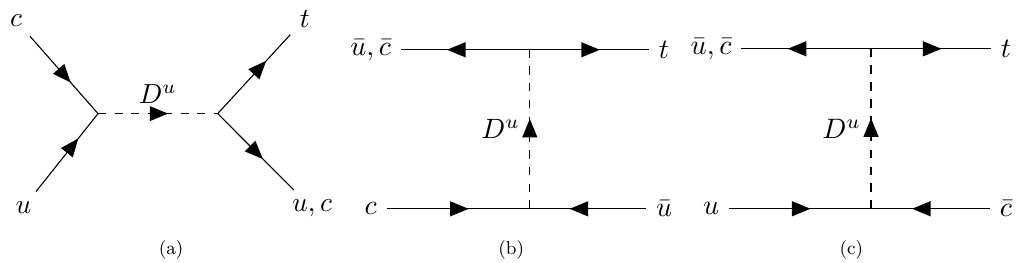}
\caption{Diquark contributions to single-top production in the
  $t$ channel in $p\,p$ collisions.}
\label{fig:diquark_stp}
\end{figure}

We now turn to single-top production (STP) \cite{Tonero:2020zcy}. In
the SM, this process occurs at LO via three modes: a $t$-channel
process $q\,b \to q'\,t$ and an $s$-channel process $q\,\overline{q}'
\to t\,\overline{b}$, both occurring through $W$-boson exchange, and a
direct $tW$ production. Of these, the dominant production mechanism at
the LHC is via the $t$-channel. Indeed, this has been measured by both
the ATLAS and the CMS Collaborations with greater precision than the
other modes. Now, the $D^u$ diquark can contribute at tree level to
this STP mode, as shown in Fig.~\ref{fig:diquark_stp}. Thus, by
comparing the measured value of the cross section for this process
with the predicted value including both the SM and NP, it is possible
to put further constraints on the mass and couplings of the $D^u$
diquark.

\begin{figure}[!t]
  \centering
  \includegraphics[width=0.75\textwidth]{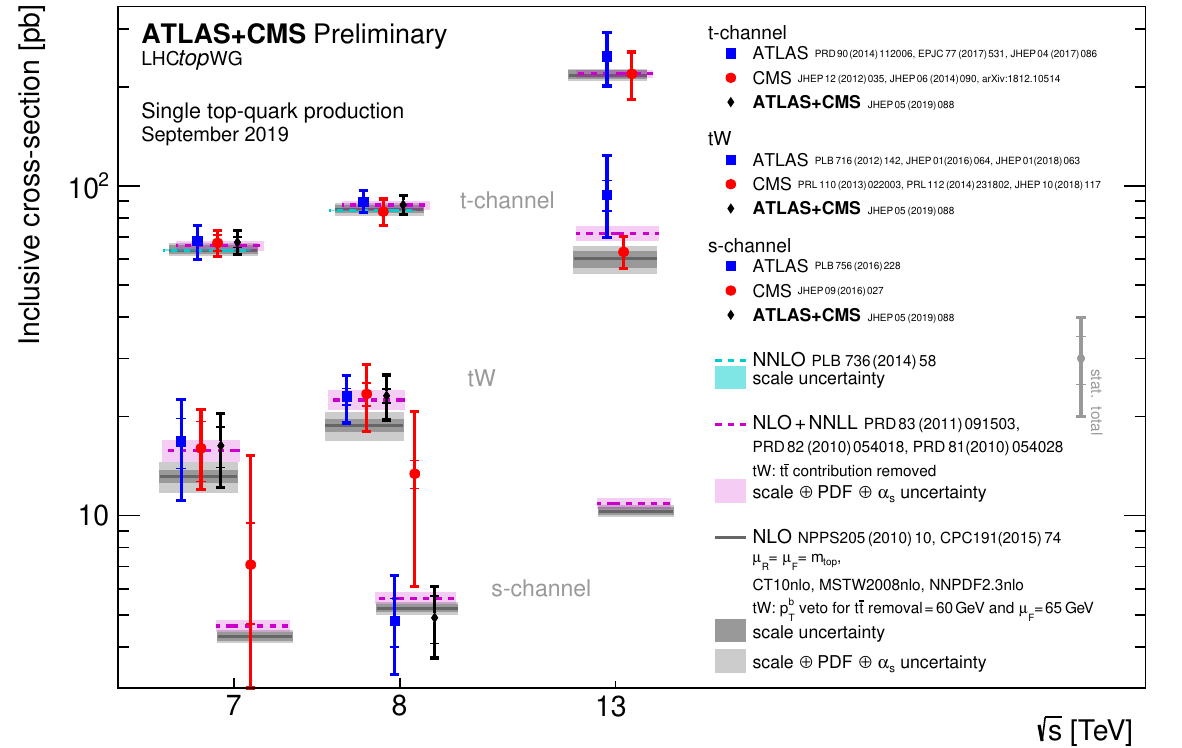}
  \caption{Summary of the available single-top production
    cross-section measurements from the LHC \cite{summary_plot}.}
  \label{fig:summary_plot}
\end{figure}

A summary of the measurements of the STP cross sections at the LHC is
shown in Fig.~\ref{fig:summary_plot}. For the combined productions of
$t$ and ${\bar t}$ quarks in the $t$ channel, we have the following
measurements: $\sigma(tq+\bar{t}q) = 89.6^{+7.1}_{-6.3}$ pb for
$\sqrt{s} = 8$ TeV from ATLAS \cite{Aaboud:2017pdi} and
$\sigma(tq+\bar{t}q) = 207 \pm 31$ pb for $\sqrt{s} = 13$ TeV from CMS
\cite{Sirunyan:2018rlu}. These must be compared to the SM prediction.
The SM prediction of the $t$-channel STP cross section for $p\,p$
collisions recommended by the ATLAS and CMS
Collaborations \cite{twiki_STP} is $\sigma_{\text{SM}}(tq+\bar{t}q)=
84.7^{+3.8}_{-3.2}$ pb at $\sqrt{s}=8$ TeV and
$\sigma_{\text{SM}}(tq+\bar{t}q)= 217^{+9.0}_{-7.7}$ pb at
$\sqrt{s}=13$ TeV. These are calculated for $m_t = 172.5$ GeV at
next-to-leading order (NLO) in QCD using Hathor v2.1
\cite{Aliev:2010zk, Kant:2014oha}. The PDF and $\alpha_S$
uncertainties are calculated using the PDF4LHC prescription
\cite{Botje:2011sn} with the MSTW2008 68\% CL NLO \cite{Martin:2009iq,
  Martin:2009bu}, CT10 NLO \cite{Lai:2010vv} and NNPDF2.3
\cite{Ball:2012cx} PDF sets, added in quadrature to the scale
uncertainty \cite{twiki_STP}.

In order to put constraints on the $D^u$ diquark, we calculate the LO
cross section $\sigma_{LO}(tq+\bar{t}q)$ for the production of a $t$
or ${\bar t}$ accompanied by a light quark from a $p\,p$ collision.
These calculations are once again performed using MadGraph5\_amc@NLO
version 2.7.2 \cite{Alwall:2014hca} by implementing the NP alongside
the SM with FeynRules \cite{Alloul:2013bka}. We use the CTEQ6L1 PDF
\cite{CTEQ6L1} and both the renormalization and factorization scales
are set to $m_t = 173\,\text{GeV}$. In order to approximate
higher-order QCD corrections, this value is then scaled with the
corresponding $K$ factors obtained from the NLO SM predictions, $K =
\sigma_{\text{NLO}}^{\text{SM}}/\sigma_{\text{LO}}^{\text{SM}}$, which
are $K=1.13$ for $\sqrt{s}=8$ TeV and $K=1.12$ for $\sqrt{s}=13$
TeV. In our analysis, we consider diquark masses of $600$ GeV and $1$
TeV, and we scan over the pair of couplings $(x^u,y^u)$, in which each
coupling varies from $0.05$ to $0.3$ in steps of $0.05$. Moreover, a
cubic interpolation is performed in order to span the region of
interest in the parameter space and to draw filled contour plots.

\begin{figure}[!t]
\centering
\includegraphics[width=\textwidth]{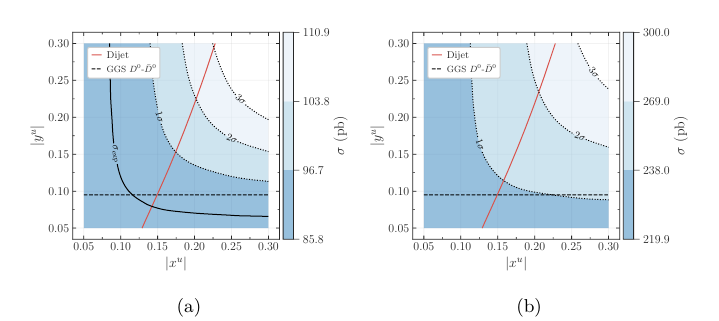}
\caption{Theoretical predictions of the total STP cross section
  $\sigma$ including the contributions of the diquark $D^u$. Results
  are shown for $M_{D^u} = 600$ GeV at (a) $\sqrt{s} = 8$ TeV and (b)
  $\sqrt{s} = 13$ TeV. The central value
  $\sigma_{\text{exp}}$ of the experimental measurement is shown 
  (black solid line), as are the $1\sigma$, $2\sigma$ and $3\sigma$ 
  regions (black dotted lines). Constraints from
  $D^0$-$\bar{D}^0$ mixing (black dashed line) \cite{GGS} and direct
  searches (red solid line) (Sec.~\ref{sec:direct_searches}) are also
  included.}
\label{fig:stp_nocuts}
\end{figure}

The predictions for the $t$-channel STP cross section at $\sqrt{s}=8$
TeV and 13 TeV, for $M_{D^u} = 600$ GeV are shown in
Fig.~\ref{fig:stp_nocuts}. The corresponding experimental measurement
is drawn, as are the boundaries of the regions describing the
$1\sigma$, $2\sigma$ and $3\sigma$ deviations from the experimental
central value. Constraints from GGS and from the direct searches
described in Sec.~\ref{sec:direct_searches} are also shown.  We see
that, at this stage, STP measurements do not lead to an improvement
over the previously-obtained constraints, \textit{i.e.,} those from
GGS and dijet measurements.  In fact, for $M_{D^u} = 1$ TeV, the
predictions for $\sigma(tq+\bar{t}q)$ lie entirely within the
$1\sigma$ region of the measurement, for all $|x^u|$ and $|y^u|$
considered.  For this reason we do not include those plots here.

In order to improve upon the earlier constraints, we focus on a
reduced phase space. To be specific, we consider two $p_T$ intervals:
$50~{\rm GeV} \leq p_T(t)\leq 300~{\rm GeV}$ and $100~{\rm GeV} \leq
p_T(t) \leq 300~{\rm GeV}$. While these cuts can be easily implemented
in MadGraph5\_amc@NLO to obtain the theory SM+NP prediction, the
corresponding measurements are not readily available.  We use
measurements of the differential cross section from ATLAS
\cite{Aaboud:2017pdi} at $\sqrt{s}$~=~8~TeV and integrate over a
subset of the bins to obtain both the experimental central value and
the uncertainties for each range of $p_T(t)$.  We obtain $47.7\pm2.5$
pb for the interval starting at $50~{\rm GeV}$ and $16.1\pm1.5$ pb for
the one starting at $100~{\rm GeV}$.  Furthermore, as our calculation
in MadGraph5\_amc@NLO is at LO, we need to obtain appropriate $K$
factors to approximate the NLO contribution. We do this by carrying
out the same integration procedure, this time on the SM prediction in
the ATLAS analysis \cite{Aaboud:2017pdi,atlas_email}.  Our LO results
are then scaled up using $K$ factors, namely $K = 1.33$ for the
interval starting at $50$ GeV and $K = 1.53$ for the interval starting
at $100$ GeV.

\begin{figure}[!t]
\centering
\includegraphics[width=\textwidth]{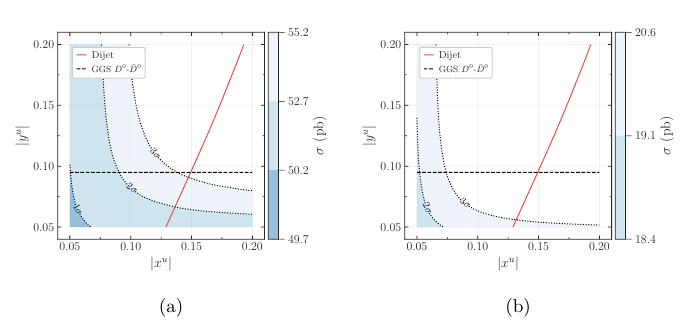}
\caption{Theoretical predictions of the total STP cross section
  $\sigma$ at $\sqrt{s} = 8$ TeV including the contributions of the
  diquark $D^u$ with $M_{D^u} = 600$ GeV. Results are shown for
  different top transverse momentum $p_T(t)$ intervals, where the
  minimum values for $p_T(t)$ are (a) $50$ GeV and (b) $100$ GeV, both
  having a maximum value of $300$ GeV. The central value
  $\sigma_{\text{exp}}$ of the experimental measurement is shown 
  (black solid line), as are the $1\sigma$, $2\sigma$ and $3\sigma$ 
  regions (black dotted lines). Constraints from $D^0$-$\bar{D}^0$ 
  mixing (black dashed line) \cite{GGS} and direct
  searches (red solid line) from Sec.~\ref{sec:direct_searches} are also
  included.}
\label{fig:stp_cuts}
\end{figure}

The theoretical predictions for the total STP cross section with
$M_{D^u} = 600$ GeV and the aforementioned cuts on $p_T(t)$ are shown
in Fig.~\ref{fig:stp_cuts}. We have varied $(x^u,y^u)$ from $0.05$ to
$0.2$ in steps of $0.01$. We now observe a reduction of the allowed
region of coupling values compared to the allowed regions from the
previously-discussed constraints. These results indicate that the
consideration of cuts can indeed strengthen the constraints on the
$D^u$ diquark.

A similar analysis can be performed for $M_{D^u} = 1$ TeV.  However,
in order to obtain an improvement on the constraints, one would have
to choose a different $p_T$ interval, excluding more of the lower
$p_T$ region and including more of the higher $p_T$ region. That is,
measurements of the differential cross section up to higher values of
$p_T$ would be required. These are as yet unavailable. If they do
become available in the future, for $\sqrt{s}$ = 8 TeV and/or 13 TeV,
it would be possible to obtain more stringent constraints on diquarks
of masses 1 TeV and higher.

\section{Conclusions}
\label{sec:conclusions}

For a variety of reasons, it is generally believed that there must be
physics beyond the SM. The clearest evidence of this NP would be if
new particles were produced at high-energy colliders.  Unfortunately,
to date, direct searches at the LHC have not found any evidence of new
particles. The other way of finding NP is through indirect searches:
if the measurement of a low-energy process disagreed with the SM
prediction, that would indicate the presence of NP. Suppose that such
an indirect signal were seen. In order to check if a particular type
of NP could be responsible, it would be necessary to (i) determine
what mass and couplings of the NP particle are required to explain the
indirect signal, and (ii) check whether such values of the mass and
couplings are consistent with constraints from direct searches. In
other words, as part of the program of indirect searches for NP, it is
important to keep track of the direct-search constraints.

In this paper, we apply this to scalar diquarks, particles that couple
to two quarks.  There are eight different types of scalar diquarks. In
Ref.~\cite{GGS}, Giudice, Gripaios and Sundrum (GGS) found that the
two most weakly-constrained diquarks are $D^u$ and $D^d$, which both
transform as a ${\mathbf{\bar 3}}$ under $SU(3)_C$, and couple
respectively to $u_{R}^{i} u_{R}^{j}$ and $d_{R}^{i} d_{R}^{j}$. To
date, these diquarks have not been observed at the LHC. We therefore
extend the GGS analysis to include the constraints from the LHC,
focusing on two masses: $M_D =$ 600 GeV and 1 TeV.

There are two types of LHC constraints. First, there are the
measurements by the CMS Collaboration of narrow dijet resonances at
$\sqrt{s} = 13$ TeV \cite{Sirunyan:2018xlo} which apply to both $D^u$
and $D^d$. We find that these measurements provide significant
improvements on the GGS constraints. Here are some examples. We denote
$x^q$ as the $D^q$ coupling to the first and second generations, and
$y^q$ as the $D^q$ couplings to the first and third or second and
third generations. For $M_D = 600$ GeV, GGS finds $|x^u| \leq 14.4$
and $|x^d y^d| \leq 0.022$ (with $|y^d| \leq 0.17$). The LHC dijet
constraints imply $|x^u| \leq 0.13$--$0.15$ and $|x^d| \leq
0.15$--$0.17$.

The second constraint applies only to $D^u$, and arises from its
indirect contribution to single top production. We find that, using
only the cross section measurements (Fig.~\ref{fig:summary_plot}),
there are no improvements on the above constraints. However, for $M_D
= 600$ GeV, we find that the allowed region of $(|x^u|,|y^u|)$
parameter space can be significantly reduced by applying a $p_T$ cut.

Finally, the reader may have wondered why we chose diquark masses of
600 GeV and 1 TeV for our detailed analysis. So far, we have not
addressed/explained this choice.  On the other hand, by now it may
have already become clear: our purpose was to demonstrate how the
constraints on the diquark parameter space can be improved using data
from dijet and STP measurements. We chose diquark masses for which
this demonstration was possible with the existing experimental
data. The LHC will run for several more years, and additional data,
particularly for STP in the high $p_T$ region, can be used in the
future to improve and extend the limits obtained here.

\bigskip
{\bf Acknowledgments}: We thank J.-F. Arguin for useful
discussions. This work was financially supported by NSERC of Canada
(BPD, DL). The work of PS was supported by the Department of Science and 
Technology, India, under the INSPIRE Faculty 
Scheme Grant No. IFA14-PH-105.

	\end{document}